\def\tilde{\widetilde}

\def\*{\star}
\def\[{\left[}
\def\]{\right]}
\def\({\left(}
\def\){\right)}

\def\frac#1#2{{#1 \over #2}}
\def\inv#1{{1 \over #1}}
\def\half{{1 \over 2}}
\def\d{\partial}

\def\ket#1{ | #1 \rangle}

\def\2pi{\hbox{$2\pi i$}}

\def\dsl{\raise.15ex\hbox{/}\kern-.57em\partial}
\def\Dsl{\,\raise.15ex\hbox{/}\mkern-.13.5mu D}

\def\th{\theta}

\def\al{\alpha}
\def\ep{\epsilon}
\def\la{\lambda}	
\def\de{\delta}		\def\De{\Delta}
		\def\Om{\Omega}

\font\numbers=cmss12
\font\upright=cmu10 scaled\magstep1
\def\stroke{\vrule height8pt width0.4pt depth-0.1pt}
\def\topfleck{\vrule height8pt width0.5pt depth-5.9pt}
\def\botfleck{\vrule height2pt width0.5pt depth0.1pt}
\def\Zmath{\vcenter{\hbox{\numbers\rlap{\rlap{Z}\kern
		0.8pt\topfleck}\kern
		2.2pt \rlap Z\kern 6pt\botfleck\kern 1pt}}}
\def\Qmath{\vcenter{\hbox{\upright\rlap{\rlap{Q}\kern
                   3.8pt\stroke}\phantom{Q}}}}
\def\Nmath{\vcenter{\hbox{\upright\rlap{I}\kern 1.7pt N}}}
\def\Cmath{\vcenter{\hbox{\upright\rlap{\rlap{C}\kern
                   3.8pt\stroke}\phantom{C}}}}
\def\Rmath{\vcenter{\hbox{\upright\rlap{I}\kern 1.7pt R}}}
\def\Z{\ifmmode\Zmath\else$\Zmath$\fi}
\def\Q{\ifmmode\Qmath\else$\Qmath$\fi}
\def\N{\ifmmode\Nmath\else$\Nmath$\fi}
\def\C{\ifmmode\Cmath\else$\Cmath$\fi}
\def\R{\ifmmode\Rmath\else$\Rmath$\fi}
\def\cadremath#1{\vbox{\hrule\hbox{\vrule\kern8pt\vbox{\kern8pt
			\hbox{$\displaystyle #1$}\kern8pt}
			\kern8pt\vrule}\hrule}}

\def\debut{ \begin{eqnarray} }
\def\fin{ \end{eqnarray} }
\def\non{ \nonumber }
\documentstyle[12pt]{article}
\title{
\hspace{4.0truein}{\small SPhT-94-043, UU-HEP/94-03}\\
\vspace{0.30truein}
{A Note on Statistical Interactions and\\ the Thermodynamic Bethe Ansatz}}
\author{Denis Bernard\footnotemark[1]  \\
  Service de Physique Th\'eorique de Saclay\footnotemark[2]\\
       F-91191, Gif-sur-Yvette, France
\\
\\
              Yong-Shi Wu\\
       Department of Physics, University of Utah\\
          Salt Lake City, Utah 84112, U.S.A. }
\footnotetext{$^*$ Member of the CNRS}
\footnotetext{$^\dagger$ \it Laboratoire de la Direction des Sciences de la
Mati\`ere du Commisariat \`a l'Energie Atomique.}
\vspace{0.2truein}
\date{March, 1994}

\begin{document}
\maketitle

\vspace{0.12truein}

\begin{abstract}
We show that the thermodynamic Bethe ansatz
equations for one-dimensional integrable
many-body systems can be reinterpreted in
such a way that they only code the
statistical interactions, in the sense of Haldane,
between particles of identical or different momenta.
Thus, the thermodynamic properties of these systems
can be characterized by the generalized ideal gases
recently proposed by one of us. For example, the
Yang-Yang $\delta$-function gas is a gas with
specific statistical interactions between
particles of different momenta, while the
Calogero-Sutherland system provides a model for an
ideal gas of particles with a fractional statistics.

\end{abstract}

\newpage
\baselineskip=20.truept

It is well-known that in 1 space dimension (or in $1+1$
space-time dimensions), bosonic theories can be mapped
into fermionic ones and reciprocally \cite{Mat,Cole}.
Furthermore, some 1d many-body systems even exhibit
a continuous boson-fermion interpolation, when
the coupling constant varies in appropriate range.
This is known to occur when the system can be solved
by the thermodynamic Bethe ansatz, such as the
Yang-Yang $\de$-function gas \cite{YY1} or the
Calogero-Sutherland system \cite{CS1}. In this note,
we give a simple characterization of
the thermodynamic properties of these systems
in terms of Haldane's statistical interactions
\cite{Ha1} and the consequential generalized
ideal gases {\cite{Wu1}.
In particular, we point out that the Bethe ansatz
equations in these models can be re-written in
such a way that all the dynamical interactions have
been transmuted into statistical interactions.

\section{Fractional statistics and thermodynamics}

Haldane \cite{Ha1} has recently introduced a notion of fractional
statistics which is independent of the dimension of space.
This notion is not based on the monodromy properties of
the $N$-particle wave functions, but on the way the
number of available single-particle states varies when
particles are added into the system. More precisely,
consider a system with a total number of particles $N=\sum_jN_j$,
with $N_j$ the number of particles of the species $j$.
Consider now adding a particle of the species $i$ into
the system without changing its size and the boundary conditions.
Keeping fixed the positions of the $N$ particles of the
original system, the wave function of the new $(N+1)$-body
system can be expanded in a basis of wave functions for the added
particle. We denote by $D_i$ the dimension of this basis.
The important point is that this dimension may depend
on the numbers $N_j$ of particles in the original system.
Assuming that this dependence is {\it linear},
Haldane defines the ``{\it statistical interaction}"
through the relation
\cite{Ha1}~:
\debut
\frac{\d D_i}{\d N_j} = - g_{ij}~. \label{eAa}
\fin
Clearly, for bosons the numbers of available single-particle states
are independent of the numbers $N_j$, and $g_{ij}=0$.
For fermions, the numbers of available single-particle states
decrease by one for each particle added, and $g_{ij}=\delta_{ij}$.

One of the ideas underlying the introduction of the generalized
Pauli principle (\ref{eAa}) is the fact that bosons and fermions can
be considered on an equal footing as far as state counting is
concerned. Indeed, for bosons or fermions, the number of states
of $(N+1)$ identical particles distributed among $G$ accessible
orbitals is~:
\debut
W_{b,f}=\frac{ (D_{b,f}+N)!}{(N+1)!\,(D_{b,f}-1)!} \non
\fin
with
\debut
D_b(N)=G,\quad {\rm and}\quad D_f(N)=G-N~. \non
\fin
One of us \cite{Wu1} has recently generalized
this to fractional statistics by directly counting
the many-body states and assuming that the total
number of states with $\{N_j\}$ particles is~:
\debut
W = \prod_i \frac{[D_i(\{N_j\})+N_i-1]!}
{N_i!\,[D_i(\{N_j\})-1]!}~,
\label{EAb}
\fin
where $D_i(\{N_j\})$ is obtained by integrating (\ref{eAa})~:
\debut
D_i(\{N_j\})+\sum_j g_{ij}\, N_j =G^0_i~, \label{EAc}
\fin
with $G^0_i\equiv D_i(\{0\})$ a constant, which is interpreted as
the number of available single-particle states when
no particle is present in the system. Namely, $G^0_i$
are the bare numbers of single-particle states.
In ref. \cite{Wu1}, $g_{ij}$ for $i\not= j$ was termed
as ``{\it mutual statistics}''.

Knowing how to enumerate states, it is then possible to study the
thermodynamics. In the thermodynamic limit, the numbers of
particles $\{N_j\}$, as well as the bare numbers of available
states $\{G^0_j\}$, become infinite. But the occupation
numbers $n_i=\({N_i/G^0_i}\)$ remain finite.
The entropy is $S=k_B\log W$ with $k_B$ the Boltzman constant.
By definition, we
call a system {\it a generalized ideal gas},
if its total energy with $(N_j)$ particles is simply given by~
\debut
E=\sum_j\ N_j\, \ep^0_j \label{EAd}
\fin
with constant $\ep^0_j$. For such gases, the thermodynamic
potential $\Om\equiv - PV$
at equilibrium can be evaluated by minimizing
\debut
\Om = E - TS - \sum_j \, N_j\mu_j \non
\fin
with respect to the variation of the densities $n_i$.
Here $T$ is the temperature and $\mu_i$ the chemical
potential for the species $i$.

The resulting thermodynamics
can be summarized as follows \cite{Wu1}~:
\debut
\Om = -k_BT\sum_i\, G^0_i\,\log\({\frac{1+w_i}{w_i}}\)~,
\label{EAf}
\fin
where the functions $w_i$ are determined by the equations~:
\debut
\log\({1+w_i}\) +\sum_j g_{ji}\,\log\({\frac{w_j}{1+w_j}}\)
= \frac{\ep^0_i-\mu_i}{k_BT}~.
\label{EAg}
\fin
These relations completely specify the thermodynamics of the
generalized ideal gas. The other thermodynamic quantities
can be derived from the relation~:
\debut
d\Om = -SdT-\sum_iN_id\mu_i-PdV~. \non
\fin
In particular, the occupation numbers $n_i$ are obtained
from $n_iG^0_i=-\d\Om/\d\mu_i$. It gives~:
\debut
n_i = \sum_j \(B^{-1}\)_{ij}~~, \label{EAh}
\fin
where $B$ is a matrix with entries~:
$B_{ij}= w_i\delta_{ij}+h_{ij}$, with $G^0_ih_{ij}=g_{ij}G^0_j$.

The simplest example considered in \cite{Wu1} is the ideal gas
of particles with a fractional statistics; i.e. $g_{ij}
=g\delta_{ij}$, $\mu_i=\mu$. In this case, the statistical
distribution of $n_i$ is then given by $n_i=1/(w_i+g)$, with
$w_i$ satisfying eq.(\ref{EAg})~ which now becomes:
\debut
w_i\, ^g(1+w_i)^{1-g} = \exp\((\ep^0_i-\mu)/k_BT\)~.
\label{AAd}
\fin
For $g=0$ (or $g=1$), we recover the bosonic (or fermionic)
occupation numbers.

\section{The Bethe ansatz equations and state counting}

The thermodynamic  Bethe ansatz (TBA) is a method  developed by
Yang and Yang \cite{YY1} for finding the thermodynamic
properties of a one-dimensional (1d) gas of particles whose
interaction is modeled by an integrable hamiltonian.
For simplicity, we will assume that the particles have no
internal quantum number.

In the Bethe ansatz approach to integrable models, the information
is encoded in the two-body $S$-matrix. We denote it by $S(k)$
with $k$ the relative momentum of the scattering particles;
we have $S(k)=- \exp(-i\th(k))$, where $\th(k)$ is the phase
shift and it is odd in $k$. The eigenstates of the $N$-body
hamiltonian (with the periodic boundary condition) are labeled
by N (pseudo-)momenta $\{k_r\}$ $(r=1,,\cdots,N)$ (defined in the
asymptotic regions), which are solutions of
the Bethe ansatz equations~:
\begin{eqnarray}
e^{ik_rL} = \prod_{s\not= r} S(k_s-k_r)\quad
{\rm for\ all}\ r~,
\label{EBa}
\end{eqnarray}
with $L$ the length of the system. The energy of the eigenstate
$\ket{k_1,\cdots,k_N}$ is found to be~:
\begin{eqnarray}
E=\sum_r \ep^0(k_r)~, \label{energ}
\end{eqnarray}
where $\ep^0(k)$ is some universal function, e.g. $\ep^0(k)=k^2$.

Using this ansatz, Yang and Yang have derived the thermodynamic
potential of such gas, which we denote by $\Om=\Om(L,T,\mu)$,
at temperature $T$ and chemical potential $\mu$ \cite{YY1}~:
\begin{eqnarray}
\frac{\Om}{L}= -k_BT\int_{-\infty}^{\infty}\frac{dk}{2\pi}
\,\log\({1+e^{-\ep(k)/k_BT}}\)~, \label{EBb}
\end{eqnarray}
where the function $\ep(k)$ is determined by solving
the TBA equations~:
\begin{eqnarray}
\ep(k)-k_BT\int_{-\infty}^{\infty}\frac{dk'}{2\pi}\,
\phi(k,k')\, \log\({1+e^{-\ep(k')/k_BT}}\) =\ep^0(k)-\mu~,
\label{EBc}
\end{eqnarray}
with $\phi(k,k')$ the derivative of the phase shift~:
\begin{eqnarray}
\phi(k,k')=i\frac{d}{dk}\log S(k,k')=\th' (k-k')~.
\label{EBe}
\end{eqnarray}

Here we point out that a simple characterization of the
statistical properties of these systems can be given in
terms of the notions we reviewed in the last section.
Namely, the thermodynamic properties of the 1d TBA gases
can be characterized as generalized ideal gases with non-trivial
statistical interactions (or mutual statistics), if
the particles with different momenta are considered as
belonging to different species (subject to one and the same
chemical potential).

As we now explain, the information about statistical
interactions is encoded just in the
Bethe ansatz equations (\ref{EBa}), if it is rewritten
in approriate form. Indeed, taking as usual the
logarithm of eq.(\ref{EBa}),  the Bethe ansatz
equations become~:
\begin{eqnarray}
L\,k_r= \sum_{s\not= r} \th(k_r-k_s) + 2\pi\, I_r~,
\label{EBf}
\end{eqnarray}
with $\th(k)$ the phase shift. Here $\{I_r\}$ $(r=1,\cdots,N)$
is a set of integers or half integers, depending on
N being odd or even, which one may choose to serve
as the quantum numbers labeling the eigenstates, instead of
the momenta $\{k_r\}$. For simplicity, let us consider the
cases when these $I_r$'s can be chosen to be all different
and still provide a complete set of solutions
(i.e., one has a fermion description for the purpose
of state counting). Though in this description the statistics
looks simple (just like fermions), the dynamics is nontrivial
in that the total energy is a complicated function of
$\{I_r\}$. Now let rewrite eq. (\ref{EBf}), in the thermodynamic
limit (with $N\to\infty$, $L\to\infty$ and $N/L$ fixed),
in the momentum description for state counting,
by dividing the range of momentum $k$ into
intervals with equal size $\De k$ and labeling each
interval by its midpoint $k_i$. (At the end of
the discussion, we will take the limit $\De k\to 0$.)
Then we treat the particles with momentum in the
$i$-th interval as belonging to the $i$-th species,
for which the bare available single-particle states
is obviously $G^0_i= L\De k/2\pi$. Now let us introduce
the distribution for the roots, $k_r$, of the Bethe
ansatz equations (\ref{EBf}) by~:
\begin{eqnarray}
L\, \rho(k_i) \De k = N_i = {\rm No.\,\, of\,\,}
k_r \,\, {\rm in\,\,}
(k_i-\De k/2, k_i+\De k/2)~.
\label{DEs}
\end{eqnarray}
Note that $2\pi \rho(k_i)=N_i/G^0_i$ is the occupation number
distribution.

Now, motivated by eq. (\ref{EBf}), let us define~:
\begin{eqnarray}
\frac{2\pi}{L}\ I(k) \equiv
k- \inv{L}\sum_{k_s}\th(k-k_s)
= k- \sum_{j} \th(k-k_j) \rho(k_j) \De k~.
\label{PRd}
\end{eqnarray}
It is easy to see that whenever $I(k)= I$ with
$I$ an integer, the corresponding value of $k$ is
a possible root of the Bethe ansatz equations
(\ref{EBf}), and thus represents an accessible
orbital. Therefore, the number of accessible orbitals,
$\tilde{D}_i(\{N_j\})\equiv L \, \rho_t(k_i) \De k$,
in the $i$-th momentum interval is given by
\debut
\tilde{D_i} = I(k_i+\De k/2)-I(k_i-\De k/2)~;
\fin
or, using eq. (\ref{PRd}) and taking the $\De k\to 0$,
we have~:
\begin{eqnarray}
\rho_t(k) =  {1\over 2\pi} - {1\over 2\pi}
\int_{-\infty}^{\infty} dk'\, \th' (k-k') \rho(k')~.
\label{SCW}
\end{eqnarray}

Recall that in the fermion description for
state counting, the same W in eq.(\ref{EAb}) is
obtained with the number of accessible orbitals
for the $i$-th species taken to be
\debut
\tilde{D_i} (\{N_j\})=D_i(\{N_j\})+N_i-1
=G^0_i+N_i-1-\sum_j g_{ij}\,N_j~.
\label{EAc1}
\fin
Comparing eq. (\ref{SCW}) with the continuum form
of this equation, we identify the statistical
interactions in the momentum description to be
\begin{eqnarray}
g(k, k')= \de (k-k')+ {1\over 2\pi}\, \th' (k-k')~.
\label{FSe}
\end{eqnarray}
This shows that the dynamical interaction, which
is summarized in the two-body phase shift, is
transmuted into a statistical interaction in the
momentum description.

The statistical interaction between particles
of identical and different momenta is a manifestation
of the following features of the spectrum of a TBA gas:
The ground state corresponds to an equidistribution of
the integral quantum numbers $\{I_r\}$ in a
certain interval: $I_{r+1}-I_r=1$; and the excited
states correspond to particle/hole excitations
in the integral lattice for $I$:
\begin{eqnarray}
I_{r+1}-I_r=1+M_r^h~, \label{eBf}
\end{eqnarray}
where $M_r^h$ is the number of holes, i.e. unoccupied
integer numbers, between $I_{r+1}$ and $I_r$. Therefore,
changing to the momentum description, we have the total
density of accessible states
\begin{eqnarray}
\rho_t(k)=\rho(k)+\rho_h(k)~, \label{eBf2}
\end{eqnarray}
where $\rho(k)$ is the particle density
defined in eq.(\ref{DEs}), and
$\rho_h(k)=M^h(k)\rho(k)$ is the hole density, which is
related to the distribution of unoccupied sites in the
integer lattice of quantum number $I$. This equation
is essentially the Bethe ansatz equation in the
thermodynamic limit. In the light of this, eq.(\ref{SCW})
implies the linear dependence of the hole density
$\rho_h (k)$ on the particle density $\rho (k')$.

Having identified $g_{ij}$ or $g(k, k')$,
we turn our attention to the condition (\ref{EAd})
for a generalized ideal gas. Because of eq. (\ref{energ}),
this condition holds good:
\begin{eqnarray}
\frac{E}{L} = \int_{-\infty}^{+\infty} dk\
\rho(k)\,\ep^0(k)~. \non
\end{eqnarray}
Thus there is no interaction energy between particles of
different momenta. Contrary to the fermion description,
in the momentum description the dynamics looks simple but
statistics is non-trivial.

Now we can apply the formulas for the generalized
ideal gas in the last section to derive the
thermodynamic properties of the TBA system.
It is not surprising that they lead to exactly the
same results as Yang-Yang's.
For example, eqs. (\ref{EAf}) and (\ref{EAg})
coincide with eqs. (\ref{EBb}) and (\ref{EBc}),
with the following idenfication:
\def\toto{ \longleftrightarrow }
\begin{eqnarray}
G^0_i/L &\toto & \rho^0(k)\equiv\inv{2\pi} \label{eBd}\\
w_i &\toto& Y(k)=\exp\({\frac{\ep(k)}{k_BT} }\) \label{eBc}\\
g_{ij} &\toto& g(k,k')= \delta(k-k')+\inv{2\pi}\phi(k,k')\label{EBd}
\end{eqnarray}
with the discrete sum replaced by the integral
over momentum. In particular, Yang-Yang's
form of the thermodynamic limit of the Bethe
ansatz equations~:
\begin{eqnarray}
\rho_h(k)+\rho(k) + \int_{-\infty}^{+\infty}\frac{dk'}{2\pi}
\,\phi(k,k')\,\rho(k') = \rho^0(k)=\inv{2\pi}
\label{SCy}
\end{eqnarray}
agrees with our state-counting equation
(\ref{EAc1}) or (\ref{SCW}), while their
entropy\cite{YY1}~:
 \begin{eqnarray}
\frac{S}{L}=\int_{-\infty}^{+\infty} dk
\[{(\rho+\rho_h)\log(\rho+\rho_h)-
\rho\log\rho- \rho_h\log\rho_h}\] \label{eentro}
\end{eqnarray}
is just $S=k_B\log W$, with $W$ given by
our eq.(\ref{EAb}), with the correspondence~:
\begin{eqnarray}
N_i/L &\toto& \rho(k) \label{EBk}\\
D_i(\{N_j\})/L &\toto& \rho_h(k)~~. \label{EBl}
\end{eqnarray}
This last idenfication is quite natural: the hole density
$\rho_h(k)$ clearly represents the density of available
states for an additional particle to be added.

\section{Examples of generalized ideal gas}

The simplest example of the TBA system is the repulsive
$\de$-function boson\footnote{Boson here specifies only
the symmetry of many-body wave functions, and does not
refer to the statistical interaction that determines or
characterizes the thermodynamic properties of the gas.}
gas, with a two-body potential: $V(x)=2c
\de (x),\, c\geq 0$ ($x$ being the relative coordinate).
This model is integrable with a fermion description
for the Bethe ansatz equations (\ref{EBf}) with the
two-body phase shift~:
\begin{eqnarray}
\th (k) = - 2\, tan^{-1}(k/c)~.
\label{EYa}
\end{eqnarray}
Therefore, $\phi(k,k')= - 2c/[(k-k')^2+c^2]$, and the
statistical interaction is~:
\begin{eqnarray}
g(k, k')= \delta (k-k') -\inv{\pi}\, {c \over c^2+(k-k')^2}~.
\label{EYb}
\end{eqnarray}
If $c\to 0$, the second term becomes $-\de (k-k')$,
so we have an ideal boson gas, as it should be.
If $c\to \infty$, the second term vanishes for all
finite $k$, and we have an ideal Fermi gas:
$g(k, k') =\de (k-k')$. For intermediate values
of $c$, the thermodynamics of the gas is that of a
generalized ideal gas with specific mutual statistics
between different momenta.

Another well-known example is provided by the
Calogero-Sutherland model \cite{CS1}.
This is a model of particles
interacting through a $1/r^2$ potential.
It is integrable, and its two-body $S$-matrix is
\begin{eqnarray}
S(k) = -\exp\[{-i\pi(\la-1)\,sgn(k)}\]~, \label{EBt}
\end{eqnarray}
with $\la$ the coupling constant. Therefore,
$\phi(k,k')=2\pi(\la-1)\delta(k-k')$, and the
statistical interaction is~:
\begin{eqnarray}
g(k,k')= \la \, \delta(k-k')~. \label{EBy}
\end{eqnarray}
The Bethe ansatz equation then reads~:
\begin{eqnarray}
\inv{2\pi} = \rho_h(k) + \la\rho(k)~. \label{EBu}
\end{eqnarray}
The coupling constant $\la$ also governs the statistical
interaction~: if the density of particles of momentum $k$
increases by a unit, then the holes density decreases
by $\la$. The bare energy is $\ep^0(k)=k^2$, and the
TBA equations reduce to algebraic equations (\ref{AAd})~:
\begin{eqnarray}
Y(k)^\la\(1+Y(k)\)^{1-\la}=\exp\((\ep^0(k)-\mu)/k_BT\)~.\non
\end{eqnarray}

Eq.(\ref{EBy}) shows that the statistical interaction is
purely between particles with identical momenta. In this
respect, the Calogero-Sutherland system appears clearly as
an ideal gas of particles with a fractional
statistics\footnote{The relation between fractional
statistics and this model has been
discussed, e.g., in ref. \cite{STat}.}, exhibiting the
phenomenon of fractional exclusion \cite{Wu1}.

Besides the values $\la=0$, which corresponds to free bosons,
the values $\la=1/2,\ 1,\ 2$ are special, since at these
values this system is related to the system of
orthogonal, unitary, symplectic random
matrices respectively. The case $\la=1$
corresponds to free fermions. The case $\la=1/2$
corresponds to ``semions" with a half statistics.
The case $\la=2$ is dual to the semionic case:
$Y^{(\la=\half)}(k;T)+1=Y^{(\la=2)}(k;-T)\,
e^{(\ep^0-\mu)/k_BT}$. In these cases, the TBA equations
can be simply solved. Finally, from eq.(\ref{EBu})
we see that the duality $\la\toto 1/\la$ \cite{Dual},
which exchanges the system with coupling
$\la$ and $1/\la$, corresponds to exchanging
particles and holes.

\section{Possible generalizations}

First let us remark that in some models a boson
description for the Bethe ansatz equations is
possible: i.e. the integers $I_r$ in
eq.(\ref{EBf}) are allowed to repeat. Our discussions
can be applied to these cases. Then the first term
in eq. (\ref{FSe}) disappears.

\bigskip
There are at least two potential generalizations of
this approach:

\noindent --- We can generalize this approach by relaxing the
assumption we made in eq. (\ref{EAd}), i.e. the energy spectrum
for each particle species is degenerate. Let us denote by
$\ep^0_i(\al_i)$ the energy spectrum with $\al_i=1,\cdots,G^0_i$,
then an alternative anstaz for the energy would be~:
\debut
E = \sum_{j,\al_j} N_j(\al_j) \ep^0_j(\al_j)~, \non
\fin
where $N_j(\al_j)$ are the occupation numbers with
$\sum_{\al_j}N_j(\al_j)=N_j$.This generalization should be
related to the non-scalar Bethe ansatz (for models with
internal degrees of freedom).

\bigskip
\noindent --- We can use this approach to generalize the
Fermi liquid theory by incorporating statistical
as well as dynamical interactions. At thermodynamic
equilibirum, the system will be described
by the particle density $\rho(\vec k)$ and the hole density
$\rho_h(\vec k)$. In $D$ space dimension, the entropy
will be assumed to be the same as in eq.(\ref{eentro})~:
\begin{eqnarray}
\frac{S}{L}=\int d^D\vec k
\[{(\rho+\rho_h)\log(\rho+\rho_h)-
\rho\log\rho- \rho_h\log\rho_h}\]~, \label{entrobis}
\end{eqnarray}
where the densities $\rho(\vec k)$ and $\rho_h(\vec k)$
are assumed to be coupled with a phenomenological
statistical interaction $g(\vec k,\vec q)$:
\debut
\rho_h(\vec k) + \int d^D\vec q\  g(\vec k,\vec q) \rho(\vec q)
= \rho^0~. \label{estati}
\fin
Mimicing the Fermi liquid theory,
the simplest ansatz for the energy is quadratic in
the particle density~:
\debut
\frac{\d E}{\d \rho(\vec k)} =  \ep^0(\vec k)
+ \int d^D\vec q\  V(\vec k,\vec q) \rho(\vec q)
\equiv \ep_{dr}(\vec k;\rho)~, \label{efermi}
\fin
where $V(\vec k,\vec q)$ is a phenomenological dynamical interaction.
As we will report elsewhere,
the thermodynamic properties can be computed from eqs.(\ref{entrobis},
\ref{estati}, \ref{efermi}).
However, since the fields creating the particles with fractional
statistics may possess fractional anomalous dimensions, we cannot
guarantee that the ansatz (\ref{efermi}) is adapted to the description
of the low-energy/long-distance behavior of a system of such particles.

Acknowledgements:
D. Bernard thanks O. Babelon for friendly discussions,
and Y.-S. Wu acknowledges helpful discussions with B. Sutherland.
Y.-S. Wu was supported in part by U.S. NSF grant PHY-9309458.

Note added: After the completion of the paper, we learn that
since last year, the problem of state counting from BAE for
the non-scalar integrable models (for particles with internal
degrees of freedom) has been studied in several papers, in some
of which the connection to the fractional statistics in
Haldane's sense was noted, though no thermodynamics was addressed.
See ref. \cite{McCoyetal}. We thank B. McCoy for telling us
the existence of these papers.

\end{document}